\documentclass[letterpaper,english,reprint,aps,prl,superscriptaddress,showpacs,longbibliography]{revtex4-2}
\usepackage[utf8]{inputenc}
\usepackage{amsmath}
\usepackage{amssymb}
\usepackage{graphicx}
\usepackage{soul}
\usepackage{color, xcolor}
\usepackage{braket}
\makeatletter

\pdfpageheight\paperheight
\pdfpagewidth\paperwidth

\usepackage[colorlinks,linkcolor=blue,anchorcolor=black,citecolor=blue,urlcolor=blue]{hyperref}

\makeatother

\usepackage{babel}

\begin{document}
	\title{Cavity-enhanced Rydberg atom microwave receiver}    
	\author{Bang Liu}
	\affiliation{Key Laboratory of Quantum Information, University of Science and Technology
		of China, Hefei, Anhui 230026, China.}
	\affiliation{Synergetic Innovation Center of Quantum Information and Quantum Physics,
		University of Science and Technology of China, Hefei, Anhui 230026,
		China.}
	\author{Li-Hua Zhang}
	\affiliation{Key Laboratory of Quantum Information, University of Science and Technology
		of China, Hefei, Anhui 230026, China.}
	\affiliation{Synergetic Innovation Center of Quantum Information and Quantum Physics,
		University of Science and Technology of China, Hefei, Anhui 230026,
		China.}
  
	\author{Zong-Kai Liu}
	\affiliation{Key Laboratory of Quantum Information, University of Science and Technology
		of China, Hefei, Anhui 230026, China.}
	\affiliation{Synergetic Innovation Center of Quantum Information and Quantum Physics,
		University of Science and Technology of China, Hefei, Anhui 230026,
		China.}

    \author{Qi-Feng Wang}
	\affiliation{Key Laboratory of Quantum Information, University of Science and Technology
		of China, Hefei, Anhui 230026, China.}
	\affiliation{Synergetic Innovation Center of Quantum Information and Quantum Physics,
		University of Science and Technology of China, Hefei, Anhui 230026,
		China.}

    \author{Yu Ma}
	\affiliation{Key Laboratory of Quantum Information, University of Science and Technology
		of China, Hefei, Anhui 230026, China.}
	\affiliation{Synergetic Innovation Center of Quantum Information and Quantum Physics,
		University of Science and Technology of China, Hefei, Anhui 230026,
		China.}

   \author{Tian-Yu Han}
	\affiliation{Key Laboratory of Quantum Information, University of Science and Technology
		of China, Hefei, Anhui 230026, China.}
	\affiliation{Synergetic Innovation Center of Quantum Information and Quantum Physics,
		University of Science and Technology of China, Hefei, Anhui 230026,
		China.}

      \author{Zheng-Yuan Zhang}
	\affiliation{Key Laboratory of Quantum Information, University of Science and Technology
		of China, Hefei, Anhui 230026, China.}
	\affiliation{Synergetic Innovation Center of Quantum Information and Quantum Physics,
		University of Science and Technology of China, Hefei, Anhui 230026,
		China.}

        \author{Shi-Yao Shao}
	\affiliation{Key Laboratory of Quantum Information, University of Science and Technology
		of China, Hefei, Anhui 230026, China.}
	\affiliation{Synergetic Innovation Center of Quantum Information and Quantum Physics,
		University of Science and Technology of China, Hefei, Anhui 230026,
		China.}

          \author{Jun Zhang}
	\affiliation{Key Laboratory of Quantum Information, University of Science and Technology
		of China, Hefei, Anhui 230026, China.}
	\affiliation{Synergetic Innovation Center of Quantum Information and Quantum Physics,
		University of Science and Technology of China, Hefei, Anhui 230026,
		China.}

          \author{Qing Li}
	\affiliation{Key Laboratory of Quantum Information, University of Science and Technology
		of China, Hefei, Anhui 230026, China.}
	\affiliation{Synergetic Innovation Center of Quantum Information and Quantum Physics,
		University of Science and Technology of China, Hefei, Anhui 230026,
		China.}

            \author{Han-Chao Chen}
	\affiliation{Key Laboratory of Quantum Information, University of Science and Technology
		of China, Hefei, Anhui 230026, China.}
	\affiliation{Synergetic Innovation Center of Quantum Information and Quantum Physics,
		University of Science and Technology of China, Hefei, Anhui 230026,
		China.}

	\author{Dong-Sheng Ding}
	\email{dds@ustc.edu.cn}
	
	\affiliation{Key Laboratory of Quantum Information, University of Science and Technology
		of China, Hefei, Anhui 230026, China.}
	\affiliation{Synergetic Innovation Center of Quantum Information and Quantum Physics,
		University of Science and Technology of China, Hefei, Anhui 230026,
		China.}
	\author{Bao-Sen Shi}
	
	\affiliation{Key Laboratory of Quantum Information, University of Science and Technology
		of China, Hefei, Anhui 230026, China.}
	\affiliation{Synergetic Innovation Center of Quantum Information and Quantum Physics,
		University of Science and Technology of China, Hefei, Anhui 230026,
		China.}
	\date{\today}

	\begin{abstract}
		Developing microwave electric field sensing based on Rydberg atom has received significant attention due to its unique advantages. However, achieving effective coupling between Rydberg atom and the microwave electric field in the sensing process is a challenging problem that greatly impacts the sensitivity. To address this, we propose the use of a microwave resonant cavity to enhance the effective coupling between the Rydberg atoms and the microwave electric field. In our experiment, we use a three-photon excitation scheme to prepare Rydberg atoms, make measurements of electric fields without and with a microwave cavity in which the vapor cell is put inside. Through experimental testing, we achieve an 18 dB enhancement of power sensitivity. The experiment shows an effective enhancement in electric field pulse signal detection. This result provides a promising direction for enhancing the sensitivity of Rydberg atomic electric field sensors and paves the way for their application in precision electric field measurements. 
	\end{abstract}

	\maketitle
	
	\section{INTRODUCTION}

 Microwave (MW) electric field sensing based on Rydberg atoms offers the ability to achieve high sensitivity as the large polarizability of Rydberg atoms \cite{Adams_2020} makes them very sensitive to external electric fields. Through the control by narrow band laser fields, its energy levels can be tailored to match the desired MW frequency range, and providing a non-destructive measurement. This flexibility allows for the development of a wide range of sensors that can operate at different frequencies \cite{Li2022continuous,simons2021continuous,PhysRevApplied.18.054003,PhysRevApplied.19.044049}, making them suitable for various applications, such as metrology \cite{9363580,10.1063/1.4984201}, sensing \cite{PhysRevA.103.063113,photonics9040250,Fan_2015}, and communication \cite{10238372,9374680,holloway2019,anderson2021}.
 
Recently, different schemes have been proposed to achieve electric field measurements based on the electromagnetically-induced transparency (EIT) effect of Rydberg atoms \cite{PhysRevLett.98.113003,sedlacek2012microwave}, and high sensitivity has been obtained with different protocols \cite{ding2022enhanced,jing2020atomic}. The rich energy level structures of the Rydberg atoms have also been used to achieve measurements of RF electric fields in the kHz-THz range \cite{below1khzJau,Waveguide2021,downes2020full,PhysRevApplied.18.014045}. The practical applications of Rydberg atomic electric field measurements have been actively explored and the great potential of Rydberg atomic electric field measurements has been demonstrated \cite{liu2022deep,PhysRevApplied.18.014033,63,42}.

However, in order to better promote the application of the Rydberg atoms in the field of MW electric field measurements, it is particularly important to improve the sensitivity of the Rydberg atoms for MW sensing. Although the quantum projection noise limit sensitivity of Rydberg atoms is theoretically advantageous compared to that of conventional antennas, it is still not able to surpass the sensitivity of conventional antennas in practice due to experimental conditions \cite{7005405}. One of the core problems is how to solve the effective coupling between the Rydberg atoms and the MW electric field \cite{PhysRevLett.124.193604,PhysRevLett.108.063004,PhysRevA.105.022626,Waveguide2021,holloway2022ssr,RN3337}. Conventional MW antennas have directional gain by collecting the MW electric field in a certain direction range for measurement, while the sensing component of the Rydberg atomic electric field measurement system is a glass vapour cell, which does not have directional gain. Additionally, the external MW electric field tends to disperse in free space, and the spatial area in which Rydberg atoms exist within the vapor cell is very limited due to constraints imposed by the excitation laser. This makes it challenging to effectively couple the Rydberg atoms with the external MW electric field, resulting in significant insertion loss and a substantial reduction in sensitivity for Rydberg atomic electric field measurements.

Here, we report an experiment of cavity enhanced Rydberg atoms MW receiver. Coupling Rydberg atoms with a MW cavity allows for strong interaction between MW radiation and Rydberg atoms, resulting in increased sensitivity to small changes of the MW signal. This can be theoretically explained by the MW resonator model, and experimentally we have achieved a 18 dB power sensitivity enhancement for the measurement of the electric field with MW cavity enhancement. This enables more accurate and precise measurements and builds the foundation for further improvement of the Rydberg atomic electric field sensing performance.
	
\section{EXPERIMENT setup}
	
    We use a Cesium atom vapour cell in our experiments to demonstrate electric field sensing. 
    \begin{figure*}[htp]
    \includegraphics[width=2\columnwidth]{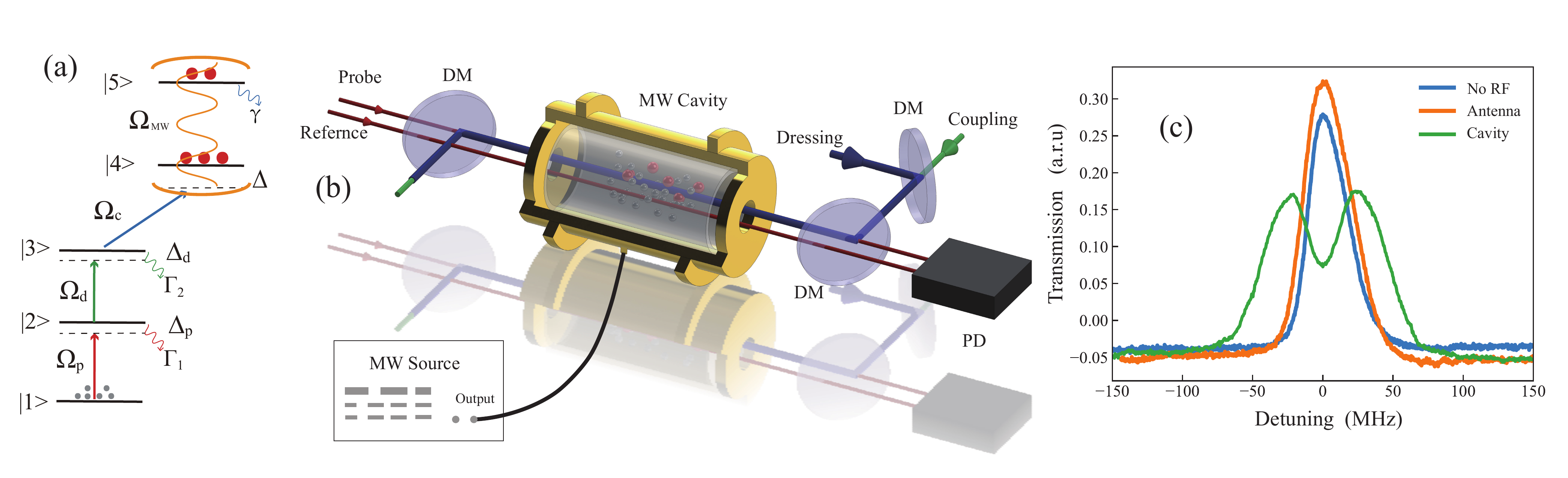}\caption{(a) Energy-level diagram of three photon Rydberg-EIT. The probe laser is resonant with transition of $\ket{1} \xrightarrow{} \ket{2}$ ($\ket{6S_{1/2}} \xrightarrow{} \ket{6P_{3/2}}$) with Rabi frequency of $\Omega_{p}$. The dressing laser drives the transition of $\ket{2} \xrightarrow{} \ket{3}$ ($\ket{6P_{3/2}} \xrightarrow{} \ket{7S_{1/2}}$) with $\Omega_{d}$. The coupling laser drives the transition of $\ket{3} \xrightarrow{} \ket{4}$ ($\ket{7S_{1/2}} \xrightarrow{} \ket{55P_{3/2}}$) with $\Omega_{c}$. The resonant frequency of the MW cavity is 4.485 GHz, which couples with electric fields at transition of $\ket{4} \xrightarrow{} \ket{5}$ ($\ket{55P_{3/2}} \xrightarrow{} \ket{54D_{5/2}}$) with $\Omega_{\rm MW}$. (b) Schematic of experimental setup. DM: dichroic mirror. PD: photodetector. (c) The EIT spectra of probe light with or without MW cavity. The blue line is the transmission of probe light without MW cavity, showing an EIT peak. The green line is the EIT spectra with MW cavity, which has two peaks caused by Autler-Townes splitting. }
    \label{fig1}
    \end{figure*}
    The energy level structure of atoms and the experimental setup are shown in Fig.~\ref{fig1}(a) and Fig.~\ref{fig1}(b). We use a three-photon excitation scheme to excite atoms from the ground state to the Rydberg state \cite{PhysRevA.100.063427}, with 852 nm probe light drives the transition of $\ket{1} \xrightarrow{} \ket{2}$ ($\ket{6S_{1/2}} \xrightarrow{} \ket{6P_{3/2}}$), 1470 nm laser drives the transition of $\ket{2} \xrightarrow{} \ket{3}$ ($\ket{6P_{3/2}} \xrightarrow{} \ket{7S_{1/2}}$), and 780 nm coupling light drives the transition of $\ket{3} \xrightarrow{} \ket{4}$ ($\ket{7S_{1/2}} \xrightarrow{} \ket{55P_{3/2}}$), corresponding to the Rabi frequencies $\Omega_{p}$, $\Omega_{d}$, $\Omega_{c}$, respectively. In the experimental setup, the probe light is divided into two beams to pass through the atomic gas cell, and the dressing light and coupling light propagate in the opposite direction to one of the probe light beams to reduce the Doppler broadening effect. Finally, the two beams are received by a balanced photoelectric detector for differential amplification measurement. 
    
    The gas cell is placed into a metal MW resonant cavity made of oxygen-free copper, and the MW field is fed through an external Sub-Miniature-A (SMA) port to drive the transition of $\ket{4} \xrightarrow{} \ket{5}$ ($\ket{55P_{3/2}} \xrightarrow{} \ket{54D_{5/2}}$) with Rabi frequency $\Omega_{\rm MW}$. At the same time, two small holes need to be opened at both ends of the resonant cavity to allow the laser to pass through. While, this leads to a decrease in the Q-factor of the resonant cavity. Efficient coupling between the Rydberg atoms and the cavity mode is essential for achieving high sensitivity. This requires careful consideration of factors such as atom-cavity detuning, cavity mode volume, and positioning of the atomic ensemble relative to the cavity field. In our setup, we strategically place the atomic cell in the center of the cavity to maximize the coupling strength.
     \begin{figure}[htp]
     \includegraphics[width=1\columnwidth]{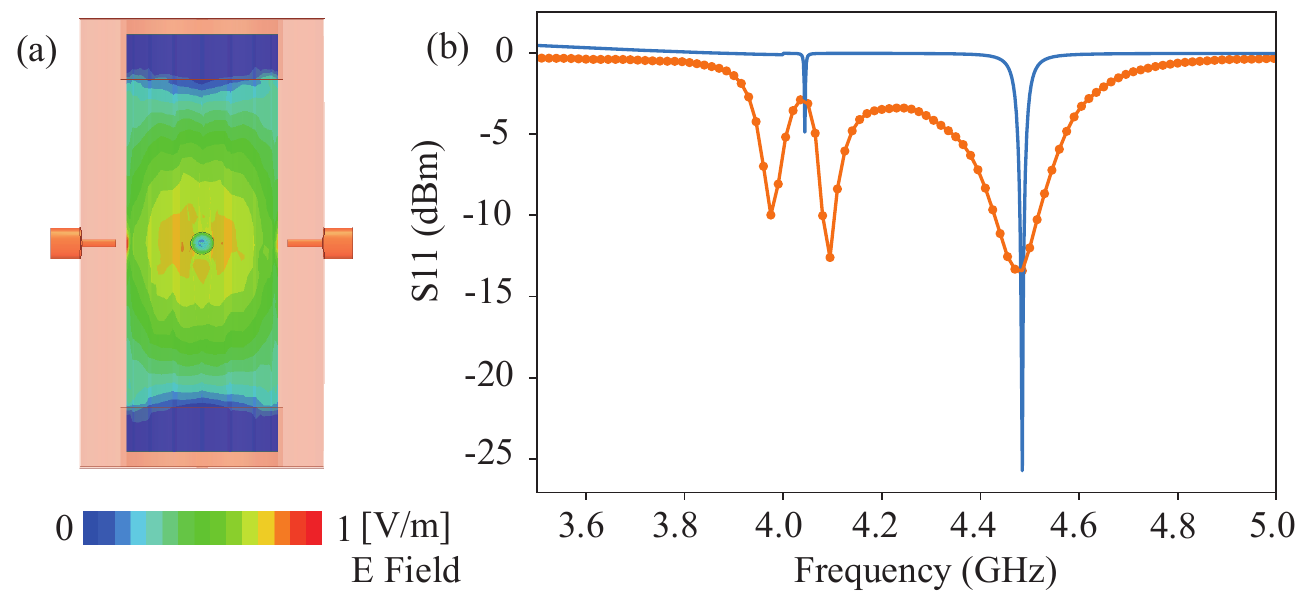}\caption{(a) Electric field distribution inside the microwave cavity. (b) $\rm S_{11}$ parameter for the microwave cavity. The blue curve is the result of a theoretical simulation, and orange data points are the results of experimental measurements.}
     \label{fig2}
     \end{figure}

   \begin{figure*}[htp]
   \includegraphics[width=2\columnwidth]{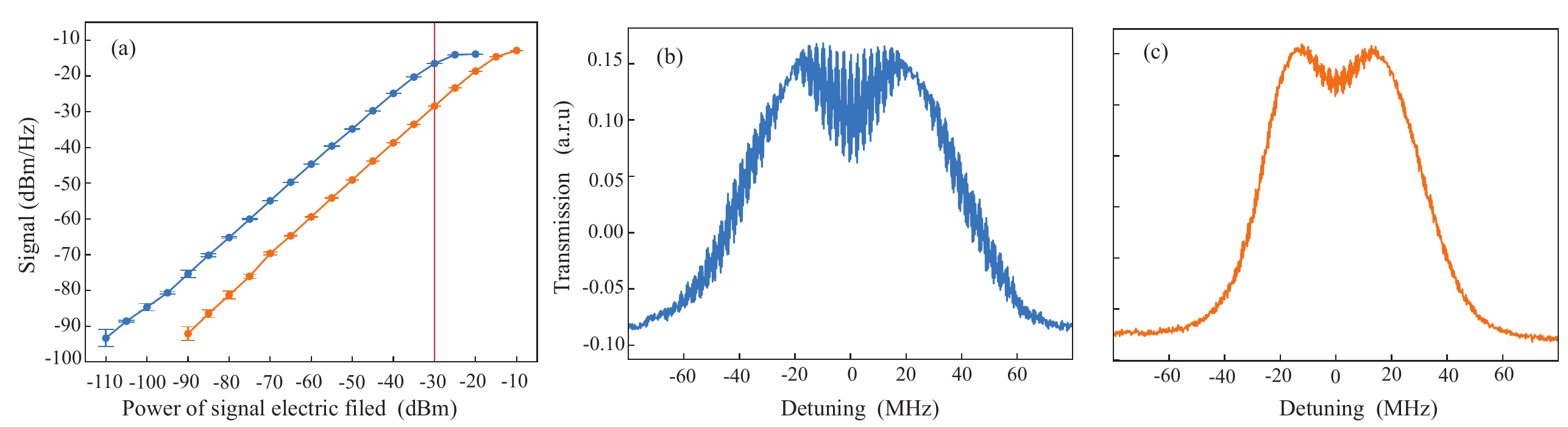}\caption{(a) Measured beat-frequency signal strength at different signal electric field powers, showing a linear response over a range of powers. The blue line is the measured data with a resonant microwave cavity, the orange one is without a cavity. The red curve marks the measured data for cavity (blue) and no-cavity (yellow) with $\rm P_{sig}=-30$ dBm signal input. (b) Measured the EIT spectra when applied a $\rm P_{sig}=-30$ dBm signal field with a cavity. (c) Measured the EIT spectra when applied a $\rm P_{sig}=-30$ dBm signal field without a cavity.}
   \label{fig3}
   \end{figure*}
	
	\section{RESULTS}

    First we simulated the MW electric field distribution in the cavity, as shown in Fig.~\ref{fig2}(a). We can observe that the field strength is at its maximum at the center of the cavity. To take advantage of this, we position the atomic gas cell inside the cavity and utilize lasers to excite the atoms to the Rydberg state precisely at the center. This arrangement ensures that the induced field strength is maximized. Additionally, we conducted a simulation of the S11 parameter for the MW cavity and compared it with the actual measurement using a vector network analyzer, as depicted in Fig.~\ref{fig2}(b). The simulation results indicate that the resonance of the MW cavity is most pronounced at 4.485 GHz, which aligns with our experimental measurements. Nevertheless, the measured S11 parameter is relatively low at -13 dB. This reduction can be attributed to both the machining process and the presence of the atomic gas cell within the cavity, both of which negatively impact the resonance performance of the MW cavity.

    After adding a MW cavity and comparing it with a general horn antenna, we investigated the EIT spectral characteristics. The results are presented in Fig.~\ref{fig1}(c). In the absence of an electric field, the EIT spectrum is represented by the blue line, exhibiting a transmission peak. When a MW electric field is applied using the horn antenna (close to the vapor cell), a slight broadening of the EIT spectrum is observed. This broadening occurs because the electric field strength is not sufficient to induce splitting in the EIT spectrum. On the other hand, when a MW resonant cavity is added and supplied with microwaves of the same power, the EIT spectrum clearly splits. This splitting is a result of the stronger coupling between the MW electric field inside the cavity and the Rydberg atoms. Moreover, the presence of the MW cavity enhances the coupling strength. To describe the coupling between the Rydberg states and the MW cavity, the Hamiltonian of the system can be expressed as:
    \begin{align*}
      \hat{H}&=\hbar\Delta_{p}\sigma_{22}+\hbar\left(\Delta_{p}+\Delta_{d}\right)\sigma_{33}+\hbar\left(\Delta_{p}+\Delta_{d}+\Delta_{c}\right)\sigma_{44} \\
      &+\hbar\left(\Delta_{p}+\Delta_{d}+\Delta_{c}+\Delta\right)\sigma_{55}-\hbar/2\left(\Omega_{p}\sigma_{12}+\Omega_{d}\sigma_{23} \right. \\
      &+\left.\Omega_{c}\sigma_{34} +\Omega_{\rm MW}\sigma_{45}+\mathrm{H.c.}\right)
    \end{align*}
where $\sigma_{i,j}=\ket{i}\bra{j}$ are atomic transition operators. In experiment, we set $ \Delta_{p}=\Delta_{d}= 0$. The Lindblad equation of system is
$$\dot{\rho} = -\frac{i}{\hbar}\left[\hat{H},\rho\right]+\sum_{ij}\mathcal{L}\left(\Gamma_{ij}\right)$$
where $\rho$ is the density matrix of system, and $\mathcal{L}\left(\Gamma_{ij}\right)=\Gamma_{ij}/2\left(2\sigma_{ji}\rho\sigma_{ij}-\sigma_{ii}\rho-\rho\sigma_{ii}\right)$ is the Lindblad operator. 

The steady-state solution of the system $ \rho_{21}\left(\Delta_{c}\right)$ can be obtained by solving the Lindblad equation, which is the EIT spectrum shown in Fig.~\ref{fig1}(c).
When in the absence of a MW electric field, the three-photon EIT phenomenon occurs, which is due to the dark state of the system. Whereas, with the MW electric field applied and ignoring the $ \Omega_{p}$, $ \Omega_{c}$, the dark state of the system splits under the MW dressing and the energy splitting is \cite{10238372,Fan_2015}
$$ \Delta E \propto \hbar \Omega_{\rm MW}$$ 
    which corresponds to the splitting of the EIT spectrum in Fig.~\ref{fig1}(c), where  $\Omega_{\rm MW}=\mu E_{\rm MW}/\hbar$ is the coupling strength of MW electric field with  Rydberg atoms.

   When an input electric field is given to the MW cavity, the circulating power in the cavity becomes larger due to the resonance effect of the cavity, making the electric field sensed by the atoms larger. When input power is $\rm P_{in}$, the power in cavity $\rm P_{c}$ is
    $$\frac{\rm P_c}{\rm P_{in}}=\left(1-\left|s_{11}\right|^2-\left|s_{21}\right|^2\right) \rm Q$$
    where $s_{11}$, $s_{21}$ are S-parameters and $\rm Q$ is the Q-factor of the  cavity. For the electric field in the cavity, we can approximate that it is uniform along the direction of the cavity and we have
    $$\rm P_{c}=\frac{1}{4}\varepsilon_0 \pi a^2 d\left|\rm E_c\right|^2$$
    where $a$ is radius of cavity, $d$ is the length of cavity and $\varepsilon_0$ is the permittivity of free space. Furthermore, the coupling strength of MW electric field inside cavity with  Rydberg atoms is $$\Omega_{\rm MW}=\frac{\mu E_{\rm MW}}{\hbar}=\frac{2\mu}{\hbar}\left[ \frac{\left(1-\left|s_{11}\right|^2-\left|s_{21}\right|^2\right) \rm P_{in} \rm Q }{\varepsilon_0 \pi a^2 d}\right]^{1/2}  $$
     So we can define the enhanced factor of the cavity  $F=\left(1-\left|s_{11}\right|^2-\left|s_{21}\right|^2\right) \rm Q$, which will enhanced the sensitivity of Rydberg atom electric field metrology. 

We exploit the resonance enhancement of the MW cavity to achieve an improved measurement of the MW. The results are presented in Fig.~\ref{fig3}. We employed the superheterodyne measurement technique to apply a strong local oscillation (LO) and a weak signal MW with a slight detuning from the LO. It was observed that the EIT spectrum is modulated by the beat frequency of the LO and the signal. The system's beat-frequency response was measured using a spectrometer, and the response intensity corresponds to the intensity of the signal under measurement. As depicted in Fig.~\ref{fig3}(a), the blue data represents the system's response with the inclusion of the MW cavity, while the orange data corresponds to the response without the cavity. It is evident that the dynamic range of both cases is approximately equal, measuring around 75 dB. However, with the addition of the MW cavity, the minimum measurable MW power can be as low as -110 dBm, which is approximately 18 dBm stronger compared to the case without the cavity. Fig.~\ref{fig3}(b) and (c) illustrate the response at an input power of -30 dBm for both scenarios.

In Fig.~\ref{fig3}(a), the blue data represents the EIT spectral features with the added MW cavity, while the orange data corresponds to the case without the cavity. It is evident that the addition of the MW cavity leads to a more pronounced modulation of the EIT spectra. This can be attributed to the enhanced MW resonance cavity. However, it should be noted that when the power of the signal electric field becomes too large, the system response deviates from linearity. This occurs due to the violation of the condition that the signal electric field should be significantly smaller than the intrinsic vibration field in the superheterodyne measurement scheme.

\begin{figure}[htp]
\includegraphics[width=1\columnwidth]{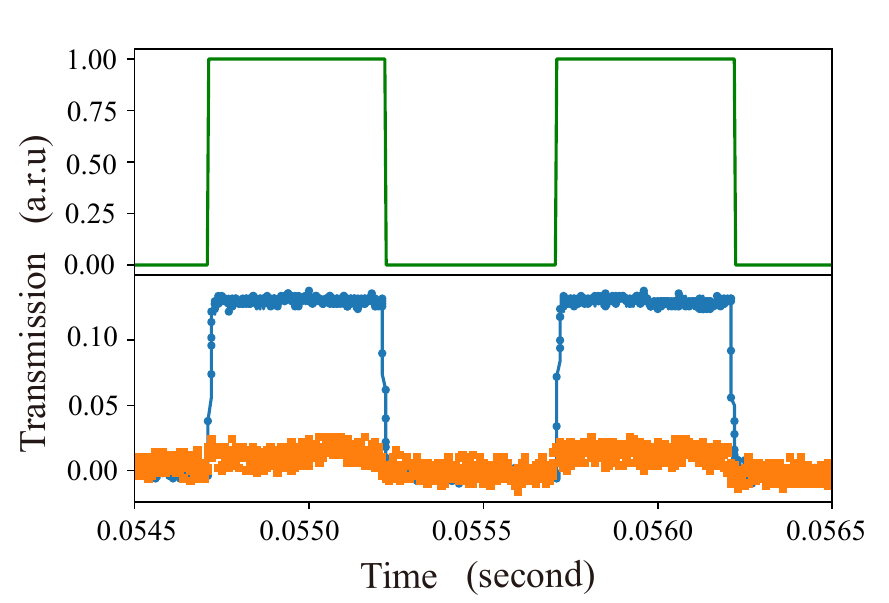}\caption{Comparison of the real-time response spectra of the Rydberg Atomic System with and without a cavity applying a square-wave modulation signal with the modulation frequency of 1 kHz, the modulation depth of 30\% and $\rm P=-30$ dBm. The green curve is the modulated signal of the carrier, the blue curve is the signal demodulated from the Rydberg atoms with the cavity, and the orange curve is the case without the cavity.}
\label{fig4}
\end{figure}

Finally, we have demonstrated the reception of the system using square-wave modulated MW signals to verify the performance of sensing MW electric fields. To do this, we applied a square-wave modulated MW signal with a modulation frequency of 1 kHz through a signal source. The signal was then received and demodulated by our Rydberg atom system, and the results obtained are shown in Fig.~\ref{fig4}. By utilizing a MW resonant cavity, we were able to achieve a significantly larger signal-to-noise ratio (blue curve) compared to the scenario without a cavity (orange curve). This enhancement in signal-to-noise ratio is particularly beneficial in the reception of weaker signals. In fact, with the addition of the cavity, the signal-to-noise ratio of the received square-wave modulated signal was boosted by a factor of $\beta=10.8$. Moreover, the use of the cavity also resulted in a reduction in the bit error rate during the reception of digitally modulated signals.
    
\section{DISCUSSION}
Due to the small volume of the laser excitation in the atomic gas cell, typically only a few hundred microns, the effective receiving area of the Rydberg atomic electric field sensor is indeed limited. This small effective area results in reduced sensitivity to the electric field being measured. The MW resonant cavity performs two primary functions. Firstly, it allows for the matching of the atoms by compressing the spatial volume of the electric field within the cavity. This compression ensures efficient interaction between the MW electric field and the Rydberg atoms. Secondly, the resonance effect of the MW cavity is utilized to enhance the interaction between the MW electric field and the Rydberg atoms. This enhancement further improves the sensitivity of the measured electric field \cite{santamariabotello2022comparison}. 

The performance of the resonant cavity plays a crucial role in this scenario. In experimental setups, a MW cavity with a relatively high Q-factor of approximately 130 is typically employed. The Q-factor can be further enhanced by reducing the cavity's volume and using low-loss dielectric materials. By achieving a Q-factor of $4\times 10^{5}$, the power sensitivity can reach as low as -180 dBm/Hz, surpassing the thermal noise limit of conventional antennas. Currently, with the use of superconducting metal surfaces as mirrors for resonant cavities, Q-factors as high as $10^{7}\sim 10^{9}$ have been achieved \cite{10.1063/1.2724816,10.1063/1.5137900}. Therefore, by combining Rydberg atoms with resonant cavities, it becomes feasible to perform measurements of extremely weak MW electric fields.

While MW cavities can significantly enhance sensitivity, it is important to note that their resonant frequency tends to be narrow. Therefore, it is crucial to ensure that the input electric field strictly matches the cavity's resonance frequency; otherwise, the performance of the cavity will be greatly compromised. In the future, there may be possibilities to broaden the operating bandwidth by adjusting the length of the MW cavity, allowing for greater flexibility in matching the resonant frequency.

\section{CONCLUSION}
 
We have successfully demonstrated an experiment of enhanced interaction between Rydberg atoms and the MW electric field by utilizing the resonance effect of the MW cavity. Through the investigation of the effective coupling between Rydberg atoms and the electric field within the cavity, we have achieved an 18 dBm increase in power sensitivity. Moreover, this enhancement effect has allowed us to improve the signal-to-noise ratio in signal demodulation, which is crucial in advancing the practical application of Rydberg atom-based electric field sensors. 

Moving forward, there are several potential avenues for further improvement. One possibility is to fabricate cavities with higher Q-factors, which would enhance the measurement sensitivity even more. Additionally, it would be worthwhile to explore the dynamics evolution of Rydberg atoms under strong coupling conditions. Furthermore, we can also investigate applications of atomic sensors in areas requiring extremely low measurement sensitivities, such as dark matter detection. Overall, our results not only contribute to the advancement of Rydberg atom-based electric field sensors but also open up new possibilities for their utilization in a range of scientific and technological applications \cite{PhysRevD.108.035042,RevModPhys.75.777,engelhardt2023detecting}.
	
	\section{List of abbreviations}
MW: Microwave
EIT: electromagnetically-induced transparency
DM: dichroic mirror
PD: photodetector
SMA: Sub-Miniature-A 
LO: local oscillation
    \section{Declarations}

\subsection{Ethical Approval and Consent to participate}
Not applicable
\subsection{Consent for publication}
Not applicable
\subsection{Availability of supporting data}
The datasets used and/or analysed during the current study are available from the corresponding author on reasonable request.
\subsection{Competing interests}
The authors declare that they have no competing interests.
\subsection{Authors' contributions}
D-S.D. proposed the project, B.L. developed the research and analysed the experimental data. B.L. performed the experiments with L-H.Z., Q-F.W., Y.M. and T-Y.H. B.L. carried out theoretical calculations. B.L. and L-H.Z. contributed to the experimental set-up. B.L. and D-S.D. wrote the manuscript. All authors contributed to discussions of the results and the manuscript and provided revisions of manuscript.
\subsection{Funding}
This research is funded by the National Key R\&D Program of China (Grant No. 2022YFA1404002), the National Natural Science Foundation of China (Grant Nos. U20A20218, 61525504, and 61435011), the Anhui Initiative in Quantum Information Technologies (Grant No. AHY020200), the major science and technology projects in Anhui Province (Grant No. 202203a13010001) and the Youth Innovation Promotion Association of the Chinese Academy of Sciences (Grant No. 2018490).
\subsection{Acknowledgments}
	
	\begin{acknowledgments}
		We are grateful to Prof. Xiao-Ming Liu from Anhui normal university for fruitful discussions about cavity designing.  
	\end{acknowledgments}

 \bibliography{ref}

\begin{thebibliography}{39}%
\makeatletter
\providecommand \@ifxundefined [1]{%
 \@ifx{#1\undefined}
}%
\providecommand \@ifnum [1]{%
 \ifnum #1\expandafter \@firstoftwo
 \else \expandafter \@secondoftwo
 \fi
}%
\providecommand \@ifx [1]{%
 \ifx #1\expandafter \@firstoftwo
 \else \expandafter \@secondoftwo
 \fi
}%
\providecommand \natexlab [1]{#1}%
\providecommand \enquote  [1]{``#1''}%
\providecommand \bibnamefont  [1]{#1}%
\providecommand \bibfnamefont [1]{#1}%
\providecommand \citenamefont [1]{#1}%
\providecommand \href@noop [0]{\@secondoftwo}%
\providecommand \href [0]{\begingroup \@sanitize@url \@href}%
\providecommand \@href[1]{\@@startlink{#1}\@@href}%
\providecommand \@@href[1]{\endgroup#1\@@endlink}%
\providecommand \@sanitize@url [0]{\catcode `\\12\catcode `\$12\catcode
  `\&12\catcode `\#12\catcode `\^12\catcode `\_12\catcode `\%12\relax}%
\providecommand \@@startlink[1]{}%
\providecommand \@@endlink[0]{}%
\providecommand \url  [0]{\begingroup\@sanitize@url \@url }%
\providecommand \@url [1]{\endgroup\@href {#1}{\urlprefix }}%
\providecommand \urlprefix  [0]{URL }%
\providecommand \Eprint [0]{\href }%
\providecommand \doibase [0]{https://doi.org/}%
\providecommand \selectlanguage [0]{\@gobble}%
\providecommand \bibinfo  [0]{\@secondoftwo}%
\providecommand \bibfield  [0]{\@secondoftwo}%
\providecommand \translation [1]{[#1]}%
\providecommand \BibitemOpen [0]{}%
\providecommand \bibitemStop [0]{}%
\providecommand \bibitemNoStop [0]{.\EOS\space}%
\providecommand \EOS [0]{\spacefactor3000\relax}%
\providecommand \BibitemShut  [1]{\csname bibitem#1\endcsname}%
\let\auto@bib@innerbib\@empty
\bibitem [{\citenamefont {Adams}\ \emph {et~al.}(2019)\citenamefont {Adams},
  \citenamefont {Pritchard},\ and\ \citenamefont {Shaffer}}]{Adams_2020}%
  \BibitemOpen
  \bibfield  {author} {\bibinfo {author} {\bibfnamefont {C.~S.}\ \bibnamefont
  {Adams}}, \bibinfo {author} {\bibfnamefont {J.~D.}\ \bibnamefont
  {Pritchard}},\ and\ \bibinfo {author} {\bibfnamefont {J.~P.}\ \bibnamefont
  {Shaffer}},\ }\bibfield  {title} {\bibinfo {title} {Rydberg atom quantum
  technologies},\ }\href {https://doi.org/10.1088/1361-6455/ab52ef} {\bibfield
  {journal} {\bibinfo  {journal} {Journal of Physics B: Atomic, Molecular and
  Optical Physics}\ }\textbf {\bibinfo {volume} {53}},\ \bibinfo {pages}
  {012002} (\bibinfo {year} {2019})}\BibitemShut {NoStop}%
\bibitem [{\citenamefont {Hu}\ \emph {et~al.}(2022)\citenamefont {Hu},
  \citenamefont {Li}, \citenamefont {Song}, \citenamefont {Bai}, \citenamefont
  {Jiao}, \citenamefont {Zhao},\ and\ \citenamefont {Jia}}]{Li2022continuous}%
  \BibitemOpen
  \bibfield  {author} {\bibinfo {author} {\bibfnamefont {J.}~\bibnamefont
  {Hu}}, \bibinfo {author} {\bibfnamefont {H.}~\bibnamefont {Li}}, \bibinfo
  {author} {\bibfnamefont {R.}~\bibnamefont {Song}}, \bibinfo {author}
  {\bibfnamefont {J.}~\bibnamefont {Bai}}, \bibinfo {author} {\bibfnamefont
  {Y.}~\bibnamefont {Jiao}}, \bibinfo {author} {\bibfnamefont {J.}~\bibnamefont
  {Zhao}},\ and\ \bibinfo {author} {\bibfnamefont {S.}~\bibnamefont {Jia}},\
  }\bibfield  {title} {\bibinfo {title} {{Continuously tunable radio frequency
  electrometry with Rydberg atoms}},\ }\href
  {https://doi.org/10.1063/5.0086357} {\bibfield  {journal} {\bibinfo
  {journal} {Applied Physics Letters}\ }\textbf {\bibinfo {volume} {121}},\
  \bibinfo {pages} {014002} (\bibinfo {year} {2022})}\BibitemShut {NoStop}%
\bibitem [{\citenamefont {Simons}\ \emph {et~al.}(2021)\citenamefont {Simons},
  \citenamefont {Artusio-Glimpse}, \citenamefont {Holloway}, \citenamefont
  {Imhof}, \citenamefont {Jefferts}, \citenamefont {Wyllie}, \citenamefont
  {Sawyer},\ and\ \citenamefont {Walker}}]{simons2021continuous}%
  \BibitemOpen
  \bibfield  {author} {\bibinfo {author} {\bibfnamefont {M.~T.}\ \bibnamefont
  {Simons}}, \bibinfo {author} {\bibfnamefont {A.~B.}\ \bibnamefont
  {Artusio-Glimpse}}, \bibinfo {author} {\bibfnamefont {C.~L.}\ \bibnamefont
  {Holloway}}, \bibinfo {author} {\bibfnamefont {E.}~\bibnamefont {Imhof}},
  \bibinfo {author} {\bibfnamefont {S.~R.}\ \bibnamefont {Jefferts}}, \bibinfo
  {author} {\bibfnamefont {R.}~\bibnamefont {Wyllie}}, \bibinfo {author}
  {\bibfnamefont {B.~C.}\ \bibnamefont {Sawyer}},\ and\ \bibinfo {author}
  {\bibfnamefont {T.~G.}\ \bibnamefont {Walker}},\ }\bibfield  {title}
  {\bibinfo {title} {Continuous radio-frequency electric-field detection
  through adjacent {R}ydberg resonance tuning},\ }\href
  {https://journals.aps.org/pra/abstract/10.1103/PhysRevA.104.032824}
  {\bibfield  {journal} {\bibinfo  {journal} {Physical Review A}\ }\textbf
  {\bibinfo {volume} {104}},\ \bibinfo {pages} {032824} (\bibinfo {year}
  {2021})}\BibitemShut {NoStop}%
\bibitem [{\citenamefont {Liu}\ \emph {et~al.}(2022{\natexlab{a}})\citenamefont
  {Liu}, \citenamefont {Liao}, \citenamefont {Zhang}, \citenamefont {Tu},
  \citenamefont {Bian}, \citenamefont {Li}, \citenamefont {Zheng},
  \citenamefont {Li}, \citenamefont {Huang}, \citenamefont {Yan},\ and\
  \citenamefont {Zhu}}]{PhysRevApplied.18.054003}%
  \BibitemOpen
  \bibfield  {author} {\bibinfo {author} {\bibfnamefont {X.-H.}\ \bibnamefont
  {Liu}}, \bibinfo {author} {\bibfnamefont {K.-Y.}\ \bibnamefont {Liao}},
  \bibinfo {author} {\bibfnamefont {Z.-X.}\ \bibnamefont {Zhang}}, \bibinfo
  {author} {\bibfnamefont {H.-T.}\ \bibnamefont {Tu}}, \bibinfo {author}
  {\bibfnamefont {W.}~\bibnamefont {Bian}}, \bibinfo {author} {\bibfnamefont
  {Z.-Q.}\ \bibnamefont {Li}}, \bibinfo {author} {\bibfnamefont {S.-Y.}\
  \bibnamefont {Zheng}}, \bibinfo {author} {\bibfnamefont {H.-H.}\ \bibnamefont
  {Li}}, \bibinfo {author} {\bibfnamefont {W.}~\bibnamefont {Huang}}, \bibinfo
  {author} {\bibfnamefont {H.}~\bibnamefont {Yan}},\ and\ \bibinfo {author}
  {\bibfnamefont {S.-L.}\ \bibnamefont {Zhu}},\ }\bibfield  {title} {\bibinfo
  {title} {Continuous-frequency microwave heterodyne detection in an atomic
  vapor cell},\ }\href {https://doi.org/10.1103/PhysRevApplied.18.054003}
  {\bibfield  {journal} {\bibinfo  {journal} {Phys. Rev. Appl.}\ }\textbf
  {\bibinfo {volume} {18}},\ \bibinfo {pages} {054003} (\bibinfo {year}
  {2022}{\natexlab{a}})}\BibitemShut {NoStop}%
\bibitem [{\citenamefont {Berweger}\ \emph {et~al.}(2023)\citenamefont
  {Berweger}, \citenamefont {Prajapati}, \citenamefont {Artusio-Glimpse},
  \citenamefont {Rotunno}, \citenamefont {Brown}, \citenamefont {Holloway},
  \citenamefont {Simons}, \citenamefont {Imhof}, \citenamefont {Jefferts},
  \citenamefont {Kayim}, \citenamefont {Viray}, \citenamefont {Wyllie},
  \citenamefont {Sawyer},\ and\ \citenamefont
  {Walker}}]{PhysRevApplied.19.044049}%
  \BibitemOpen
  \bibfield  {author} {\bibinfo {author} {\bibfnamefont {S.}~\bibnamefont
  {Berweger}}, \bibinfo {author} {\bibfnamefont {N.}~\bibnamefont {Prajapati}},
  \bibinfo {author} {\bibfnamefont {A.~B.}\ \bibnamefont {Artusio-Glimpse}},
  \bibinfo {author} {\bibfnamefont {A.~P.}\ \bibnamefont {Rotunno}}, \bibinfo
  {author} {\bibfnamefont {R.}~\bibnamefont {Brown}}, \bibinfo {author}
  {\bibfnamefont {C.~L.}\ \bibnamefont {Holloway}}, \bibinfo {author}
  {\bibfnamefont {M.~T.}\ \bibnamefont {Simons}}, \bibinfo {author}
  {\bibfnamefont {E.}~\bibnamefont {Imhof}}, \bibinfo {author} {\bibfnamefont
  {S.~R.}\ \bibnamefont {Jefferts}}, \bibinfo {author} {\bibfnamefont {B.~N.}\
  \bibnamefont {Kayim}}, \bibinfo {author} {\bibfnamefont {M.~A.}\ \bibnamefont
  {Viray}}, \bibinfo {author} {\bibfnamefont {R.}~\bibnamefont {Wyllie}},
  \bibinfo {author} {\bibfnamefont {B.~C.}\ \bibnamefont {Sawyer}},\ and\
  \bibinfo {author} {\bibfnamefont {T.~G.}\ \bibnamefont {Walker}},\ }\bibfield
   {title} {\bibinfo {title} {Rydberg-state engineering: Investigations of
  tuning schemes for continuous frequency sensing},\ }\href
  {https://doi.org/10.1103/PhysRevApplied.19.044049} {\bibfield  {journal}
  {\bibinfo  {journal} {Phys. Rev. Appl.}\ }\textbf {\bibinfo {volume} {19}},\
  \bibinfo {pages} {044049} (\bibinfo {year} {2023})}\BibitemShut {NoStop}%
\bibitem [{\citenamefont {Anderson}\ \emph
  {et~al.}(2021{\natexlab{a}})\citenamefont {Anderson}, \citenamefont
  {Sapiro},\ and\ \citenamefont {Raithel}}]{9363580}%
  \BibitemOpen
  \bibfield  {author} {\bibinfo {author} {\bibfnamefont {D.~A.}\ \bibnamefont
  {Anderson}}, \bibinfo {author} {\bibfnamefont {R.~E.}\ \bibnamefont
  {Sapiro}},\ and\ \bibinfo {author} {\bibfnamefont {G.}~\bibnamefont
  {Raithel}},\ }\bibfield  {title} {\bibinfo {title} {A self-calibrated
  si-traceable {Rydberg} atom-based radio frequency electric field probe and
  measurement instrument},\ }\href {https://doi.org/10.1109/TAP.2021.3060540}
  {\bibfield  {journal} {\bibinfo  {journal} {IEEE Transactions on Antennas and
  Propagation}\ }\textbf {\bibinfo {volume} {69}},\ \bibinfo {pages} {5931}
  (\bibinfo {year} {2021}{\natexlab{a}})}\BibitemShut {NoStop}%
\bibitem [{\citenamefont {Holloway}\ \emph {et~al.}(2017)\citenamefont
  {Holloway}, \citenamefont {Simons}, \citenamefont {Gordon}, \citenamefont
  {Dienstfrey}, \citenamefont {Anderson},\ and\ \citenamefont
  {Raithel}}]{10.1063/1.4984201}%
  \BibitemOpen
  \bibfield  {author} {\bibinfo {author} {\bibfnamefont {C.~L.}\ \bibnamefont
  {Holloway}}, \bibinfo {author} {\bibfnamefont {M.~T.}\ \bibnamefont
  {Simons}}, \bibinfo {author} {\bibfnamefont {J.~A.}\ \bibnamefont {Gordon}},
  \bibinfo {author} {\bibfnamefont {A.}~\bibnamefont {Dienstfrey}}, \bibinfo
  {author} {\bibfnamefont {D.~A.}\ \bibnamefont {Anderson}},\ and\ \bibinfo
  {author} {\bibfnamefont {G.}~\bibnamefont {Raithel}},\ }\bibfield  {title}
  {\bibinfo {title} {{Electric field metrology for SI traceability: Systematic
  measurement uncertainties in electromagnetically induced transparency in
  atomic vapor}},\ }\href {https://doi.org/10.1063/1.4984201} {\bibfield
  {journal} {\bibinfo  {journal} {Journal of Applied Physics}\ }\textbf
  {\bibinfo {volume} {121}},\ \bibinfo {pages} {233106} (\bibinfo {year}
  {2017})}\BibitemShut {NoStop}%
\bibitem [{\citenamefont {Jia}\ \emph {et~al.}(2021)\citenamefont {Jia},
  \citenamefont {Liu}, \citenamefont {Mei}, \citenamefont {Yu}, \citenamefont
  {Zhang}, \citenamefont {Lin}, \citenamefont {Dong}, \citenamefont {Zhang},
  \citenamefont {Xie},\ and\ \citenamefont {Zhong}}]{PhysRevA.103.063113}%
  \BibitemOpen
  \bibfield  {author} {\bibinfo {author} {\bibfnamefont {F.-D.}\ \bibnamefont
  {Jia}}, \bibinfo {author} {\bibfnamefont {X.-B.}\ \bibnamefont {Liu}},
  \bibinfo {author} {\bibfnamefont {J.}~\bibnamefont {Mei}}, \bibinfo {author}
  {\bibfnamefont {Y.-H.}\ \bibnamefont {Yu}}, \bibinfo {author} {\bibfnamefont
  {H.-Y.}\ \bibnamefont {Zhang}}, \bibinfo {author} {\bibfnamefont {Z.-Q.}\
  \bibnamefont {Lin}}, \bibinfo {author} {\bibfnamefont {H.-Y.}\ \bibnamefont
  {Dong}}, \bibinfo {author} {\bibfnamefont {J.}~\bibnamefont {Zhang}},
  \bibinfo {author} {\bibfnamefont {F.}~\bibnamefont {Xie}},\ and\ \bibinfo
  {author} {\bibfnamefont {Z.-P.}\ \bibnamefont {Zhong}},\ }\bibfield  {title}
  {\bibinfo {title} {Span shift and extension of quantum microwave electrometry
  with {Rydberg} atoms dressed by an auxiliary microwave field},\ }\href
  {https://doi.org/10.1103/PhysRevA.103.063113} {\bibfield  {journal} {\bibinfo
   {journal} {Phys. Rev. A}\ }\textbf {\bibinfo {volume} {103}},\ \bibinfo
  {pages} {063113} (\bibinfo {year} {2021})}\BibitemShut {NoStop}%
\bibitem [{\citenamefont {Cai}\ \emph {et~al.}(2022)\citenamefont {Cai},
  \citenamefont {Xu}, \citenamefont {You},\ and\ \citenamefont
  {Liu}}]{photonics9040250}%
  \BibitemOpen
  \bibfield  {author} {\bibinfo {author} {\bibfnamefont {M.}~\bibnamefont
  {Cai}}, \bibinfo {author} {\bibfnamefont {Z.}~\bibnamefont {Xu}}, \bibinfo
  {author} {\bibfnamefont {S.}~\bibnamefont {You}},\ and\ \bibinfo {author}
  {\bibfnamefont {H.}~\bibnamefont {Liu}},\ }\bibfield  {title} {\bibinfo
  {title} {Sensitivity improvement and determination of {R}ydberg atom-based
  microwave sensor},\ }\href {https://www.mdpi.com/2304-6732/9/4/250}
  {\bibfield  {journal} {\bibinfo  {journal} {Photonics}\ }\textbf {\bibinfo
  {volume} {9}} (\bibinfo {year} {2022})}\BibitemShut {NoStop}%
\bibitem [{\citenamefont {Fan}\ \emph {et~al.}(2015)\citenamefont {Fan},
  \citenamefont {Kumar}, \citenamefont {Sedlacek}, \citenamefont {Kübler},
  \citenamefont {Karimkashi},\ and\ \citenamefont {Shaffer}}]{Fan_2015}%
  \BibitemOpen
  \bibfield  {author} {\bibinfo {author} {\bibfnamefont {H.}~\bibnamefont
  {Fan}}, \bibinfo {author} {\bibfnamefont {S.}~\bibnamefont {Kumar}}, \bibinfo
  {author} {\bibfnamefont {J.}~\bibnamefont {Sedlacek}}, \bibinfo {author}
  {\bibfnamefont {H.}~\bibnamefont {Kübler}}, \bibinfo {author} {\bibfnamefont
  {S.}~\bibnamefont {Karimkashi}},\ and\ \bibinfo {author} {\bibfnamefont
  {J.~P.}\ \bibnamefont {Shaffer}},\ }\bibfield  {title} {\bibinfo {title}
  {Atom based rf electric field sensing},\ }\href
  {https://doi.org/10.1088/0953-4075/48/20/202001} {\bibfield  {journal}
  {\bibinfo  {journal} {Journal of Physics B: Atomic, Molecular and Optical
  Physics}\ }\textbf {\bibinfo {volume} {48}},\ \bibinfo {pages} {202001}
  (\bibinfo {year} {2015})}\BibitemShut {NoStop}%
\bibitem [{\citenamefont {Liu}\ \emph {et~al.}(2023)\citenamefont {Liu},
  \citenamefont {Zhang}, \citenamefont {Liu}, \citenamefont {Deng},
  \citenamefont {Ding}, \citenamefont {Shi},\ and\ \citenamefont
  {Guo}}]{10238372}%
  \BibitemOpen
  \bibfield  {author} {\bibinfo {author} {\bibfnamefont {B.}~\bibnamefont
  {Liu}}, \bibinfo {author} {\bibfnamefont {L.}~\bibnamefont {Zhang}}, \bibinfo
  {author} {\bibfnamefont {Z.}~\bibnamefont {Liu}}, \bibinfo {author}
  {\bibfnamefont {Z.}~\bibnamefont {Deng}}, \bibinfo {author} {\bibfnamefont
  {D.}~\bibnamefont {Ding}}, \bibinfo {author} {\bibfnamefont {B.}~\bibnamefont
  {Shi}},\ and\ \bibinfo {author} {\bibfnamefont {G.}~\bibnamefont {Guo}},\
  }\bibfield  {title} {\bibinfo {title} {Electric field measurement and
  application based on {R}ydberg atoms},\ }\href
  {https://doi.org/10.23919/emsci.2022.0015} {\bibfield  {journal} {\bibinfo
  {journal} {Electromagnetic Science}\ }\textbf {\bibinfo {volume} {1}},\
  \bibinfo {pages} {1} (\bibinfo {year} {2023})}\BibitemShut {NoStop}%
\bibitem [{\citenamefont {Fancher}\ \emph {et~al.}(2021)\citenamefont
  {Fancher}, \citenamefont {Scherer}, \citenamefont {John},\ and\ \citenamefont
  {Marlow}}]{9374680}%
  \BibitemOpen
  \bibfield  {author} {\bibinfo {author} {\bibfnamefont {C.~T.}\ \bibnamefont
  {Fancher}}, \bibinfo {author} {\bibfnamefont {D.~R.}\ \bibnamefont
  {Scherer}}, \bibinfo {author} {\bibfnamefont {M.~C.~S.}\ \bibnamefont
  {John}},\ and\ \bibinfo {author} {\bibfnamefont {B.~L.~S.}\ \bibnamefont
  {Marlow}},\ }\bibfield  {title} {\bibinfo {title} {{R}ydberg atom electric
  field sensors for communications and sensing},\ }\href
  {https://ieeexplore.ieee.org/document/9374680/} {\bibfield  {journal}
  {\bibinfo  {journal} {IEEE Transactions on Quantum Engineering}\ }\textbf
  {\bibinfo {volume} {2}},\ \bibinfo {pages} {1} (\bibinfo {year}
  {2021})}\BibitemShut {NoStop}%
\bibitem [{\citenamefont {Holloway}\ \emph {et~al.}(2021)\citenamefont
  {Holloway}, \citenamefont {Simons}, \citenamefont {Haddab}, \citenamefont
  {Gordon}, \citenamefont {Anderson}, \citenamefont {Raithel},\ and\
  \citenamefont {Voran}}]{holloway2019}%
  \BibitemOpen
  \bibfield  {author} {\bibinfo {author} {\bibfnamefont {C.}~\bibnamefont
  {Holloway}}, \bibinfo {author} {\bibfnamefont {M.}~\bibnamefont {Simons}},
  \bibinfo {author} {\bibfnamefont {A.~H.}\ \bibnamefont {Haddab}}, \bibinfo
  {author} {\bibfnamefont {J.~A.}\ \bibnamefont {Gordon}}, \bibinfo {author}
  {\bibfnamefont {D.~A.}\ \bibnamefont {Anderson}}, \bibinfo {author}
  {\bibfnamefont {G.}~\bibnamefont {Raithel}},\ and\ \bibinfo {author}
  {\bibfnamefont {S.}~\bibnamefont {Voran}},\ }\bibfield  {title} {\bibinfo
  {title} {A multiple-band {R}ydberg atom-based receiver: Am/fm stereo
  reception},\ }\href {https://doi.org/10.1109/MAP.2020.2976914} {\bibfield
  {journal} {\bibinfo  {journal} {IEEE Antennas and Propagation Magazine}\
  }\textbf {\bibinfo {volume} {63}},\ \bibinfo {pages} {63} (\bibinfo {year}
  {2021})}\BibitemShut {NoStop}%
\bibitem [{\citenamefont {Anderson}\ \emph
  {et~al.}(2021{\natexlab{b}})\citenamefont {Anderson}, \citenamefont
  {Sapiro},\ and\ \citenamefont {Raithel}}]{anderson2021}%
  \BibitemOpen
  \bibfield  {author} {\bibinfo {author} {\bibfnamefont {D.~A.}\ \bibnamefont
  {Anderson}}, \bibinfo {author} {\bibfnamefont {R.~E.}\ \bibnamefont
  {Sapiro}},\ and\ \bibinfo {author} {\bibfnamefont {G.}~\bibnamefont
  {Raithel}},\ }\bibfield  {title} {\bibinfo {title} {An atomic receiver for am
  and fm radio communication},\ }\href
  {https://doi.org/10.1109/TAP.2020.2987112} {\bibfield  {journal} {\bibinfo
  {journal} {IEEE Transactions on Antennas and Propagation}\ }\textbf {\bibinfo
  {volume} {69}},\ \bibinfo {pages} {2455} (\bibinfo {year}
  {2021}{\natexlab{b}})}\BibitemShut {NoStop}%
\bibitem [{\citenamefont {Mohapatra}\ \emph {et~al.}(2007)\citenamefont
  {Mohapatra}, \citenamefont {Jackson},\ and\ \citenamefont
  {Adams}}]{PhysRevLett.98.113003}%
  \BibitemOpen
  \bibfield  {author} {\bibinfo {author} {\bibfnamefont {A.~K.}\ \bibnamefont
  {Mohapatra}}, \bibinfo {author} {\bibfnamefont {T.~R.}\ \bibnamefont
  {Jackson}},\ and\ \bibinfo {author} {\bibfnamefont {C.~S.}\ \bibnamefont
  {Adams}},\ }\bibfield  {title} {\bibinfo {title} {Coherent optical detection
  of highly excited {Rydberg} states using electromagnetically induced
  transparency},\ }\href {https://doi.org/10.1103/PhysRevLett.98.113003}
  {\bibfield  {journal} {\bibinfo  {journal} {Phys. Rev. Lett.}\ }\textbf
  {\bibinfo {volume} {98}},\ \bibinfo {pages} {113003} (\bibinfo {year}
  {2007})}\BibitemShut {NoStop}%
\bibitem [{\citenamefont {Sedlacek}\ \emph {et~al.}(2012)\citenamefont
  {Sedlacek}, \citenamefont {Schwettmann}, \citenamefont {K{\"u}bler},
  \citenamefont {L{\"o}w}, \citenamefont {Pfau},\ and\ \citenamefont
  {Shaffer}}]{sedlacek2012microwave}%
  \BibitemOpen
  \bibfield  {author} {\bibinfo {author} {\bibfnamefont {J.~A.}\ \bibnamefont
  {Sedlacek}}, \bibinfo {author} {\bibfnamefont {A.}~\bibnamefont
  {Schwettmann}}, \bibinfo {author} {\bibfnamefont {H.}~\bibnamefont
  {K{\"u}bler}}, \bibinfo {author} {\bibfnamefont {R.}~\bibnamefont {L{\"o}w}},
  \bibinfo {author} {\bibfnamefont {T.}~\bibnamefont {Pfau}},\ and\ \bibinfo
  {author} {\bibfnamefont {J.~P.}\ \bibnamefont {Shaffer}},\ }\bibfield
  {title} {\bibinfo {title} {Microwave electrometry with {R}ydberg atoms in a
  vapour cell using bright atomic resonances},\ }\href
  {https://www.nature.com/articles/nphys2423} {\bibfield  {journal} {\bibinfo
  {journal} {Nat.Phys.}\ }\textbf {\bibinfo {volume} {8}},\ \bibinfo {pages}
  {819} (\bibinfo {year} {2012})}\BibitemShut {NoStop}%
\bibitem [{\citenamefont {Ding}\ \emph {et~al.}(2022)\citenamefont {Ding},
  \citenamefont {Liu}, \citenamefont {Shi}, \citenamefont {Guo}, \citenamefont
  {M{\o}lmer},\ and\ \citenamefont {Adams}}]{ding2022enhanced}%
  \BibitemOpen
  \bibfield  {author} {\bibinfo {author} {\bibfnamefont {D.-S.}\ \bibnamefont
  {Ding}}, \bibinfo {author} {\bibfnamefont {Z.-K.}\ \bibnamefont {Liu}},
  \bibinfo {author} {\bibfnamefont {B.-S.}\ \bibnamefont {Shi}}, \bibinfo
  {author} {\bibfnamefont {G.-C.}\ \bibnamefont {Guo}}, \bibinfo {author}
  {\bibfnamefont {K.}~\bibnamefont {M{\o}lmer}},\ and\ \bibinfo {author}
  {\bibfnamefont {C.~S.}\ \bibnamefont {Adams}},\ }\bibfield  {title} {\bibinfo
  {title} {Enhanced metrology at the critical point of a many-body {R}ydberg
  atomic system},\ }\href {https://www.nature.com/articles/s41567-022-01777-8}
  {\bibfield  {journal} {\bibinfo  {journal} {Nature Physics}\ }\textbf
  {\bibinfo {volume} {18}},\ \bibinfo {pages} {1447} (\bibinfo {year}
  {2022})}\BibitemShut {NoStop}%
\bibitem [{\citenamefont {Jing}\ \emph {et~al.}(2020)\citenamefont {Jing},
  \citenamefont {Hu}, \citenamefont {Ma}, \citenamefont {Zhang}, \citenamefont
  {Zhang}, \citenamefont {Xiao},\ and\ \citenamefont {Jia}}]{jing2020atomic}%
  \BibitemOpen
  \bibfield  {author} {\bibinfo {author} {\bibfnamefont {M.}~\bibnamefont
  {Jing}}, \bibinfo {author} {\bibfnamefont {Y.}~\bibnamefont {Hu}}, \bibinfo
  {author} {\bibfnamefont {J.}~\bibnamefont {Ma}}, \bibinfo {author}
  {\bibfnamefont {H.}~\bibnamefont {Zhang}}, \bibinfo {author} {\bibfnamefont
  {L.}~\bibnamefont {Zhang}}, \bibinfo {author} {\bibfnamefont
  {L.}~\bibnamefont {Xiao}},\ and\ \bibinfo {author} {\bibfnamefont
  {S.}~\bibnamefont {Jia}},\ }\bibfield  {title} {\bibinfo {title} {Atomic
  superheterodyne receiver based on microwave-dressed {R}ydberg spectroscopy},\
  }\href {https://www.nature.com/articles/s41567-020-0918-5} {\bibfield
  {journal} {\bibinfo  {journal} {Nat. Phys.}\ }\textbf {\bibinfo {volume}
  {16}},\ \bibinfo {pages} {911} (\bibinfo {year} {2020})}\BibitemShut
  {NoStop}%
\bibitem [{\citenamefont {Jau}\ and\ \citenamefont
  {Carter}(2020)}]{below1khzJau}%
  \BibitemOpen
  \bibfield  {author} {\bibinfo {author} {\bibfnamefont {Y.-Y.}\ \bibnamefont
  {Jau}}\ and\ \bibinfo {author} {\bibfnamefont {T.}~\bibnamefont {Carter}},\
  }\bibfield  {title} {\bibinfo {title} {Vapor-cell-based atomic electrometry
  for detection frequencies below 1 k{H}z},\ }\href
  {https://link.aps.org/doi/10.1103/PhysRevApplied.13.054034} {\bibfield
  {journal} {\bibinfo  {journal} {Phys. Rev. Appl.}\ }\textbf {\bibinfo
  {volume} {13}},\ \bibinfo {pages} {054034} (\bibinfo {year}
  {2020})}\BibitemShut {NoStop}%
\bibitem [{\citenamefont {Meyer}\ \emph {et~al.}(2021)\citenamefont {Meyer},
  \citenamefont {Kunz},\ and\ \citenamefont {Cox}}]{Waveguide2021}%
  \BibitemOpen
  \bibfield  {author} {\bibinfo {author} {\bibfnamefont {D.~H.}\ \bibnamefont
  {Meyer}}, \bibinfo {author} {\bibfnamefont {P.~D.}\ \bibnamefont {Kunz}},\
  and\ \bibinfo {author} {\bibfnamefont {K.~C.}\ \bibnamefont {Cox}},\
  }\bibfield  {title} {\bibinfo {title} {Waveguide-coupled {R}ydberg spectrum
  analyzer from 0 to 20 {GH}z},\ }\href
  {https://link.aps.org/doi/10.1103/PhysRevApplied.15.014053} {\bibfield
  {journal} {\bibinfo  {journal} {Phys. Rev. Applied}\ }\textbf {\bibinfo
  {volume} {15}},\ \bibinfo {pages} {014053} (\bibinfo {year}
  {2021})}\BibitemShut {NoStop}%
\bibitem [{\citenamefont {Downes}\ \emph {et~al.}(2020)\citenamefont {Downes},
  \citenamefont {MacKellar}, \citenamefont {Whiting}, \citenamefont
  {Bourgenot}, \citenamefont {Adams},\ and\ \citenamefont
  {Weatherill}}]{downes2020full}%
  \BibitemOpen
  \bibfield  {author} {\bibinfo {author} {\bibfnamefont {L.~A.}\ \bibnamefont
  {Downes}}, \bibinfo {author} {\bibfnamefont {A.~R.}\ \bibnamefont
  {MacKellar}}, \bibinfo {author} {\bibfnamefont {D.~J.}\ \bibnamefont
  {Whiting}}, \bibinfo {author} {\bibfnamefont {C.}~\bibnamefont {Bourgenot}},
  \bibinfo {author} {\bibfnamefont {C.~S.}\ \bibnamefont {Adams}},\ and\
  \bibinfo {author} {\bibfnamefont {K.~J.}\ \bibnamefont {Weatherill}},\
  }\bibfield  {title} {\bibinfo {title} {Full-field terahertz imaging at
  kilohertz frame rates using atomic vapor},\ }\href
  {https://doi.org/10.1103/PhysRevX.10.011027} {\bibfield  {journal} {\bibinfo
  {journal} {Phys. Rev. X}\ }\textbf {\bibinfo {volume} {10}},\ \bibinfo
  {pages} {011027} (\bibinfo {year} {2020})}\BibitemShut {NoStop}%
\bibitem [{\citenamefont {Liu}\ \emph {et~al.}(2022{\natexlab{b}})\citenamefont
  {Liu}, \citenamefont {Zhang}, \citenamefont {Liu}, \citenamefont {Zhang},
  \citenamefont {Zhu}, \citenamefont {Gao}, \citenamefont {Guo}, \citenamefont
  {Ding},\ and\ \citenamefont {Shi}}]{PhysRevApplied.18.014045}%
  \BibitemOpen
  \bibfield  {author} {\bibinfo {author} {\bibfnamefont {B.}~\bibnamefont
  {Liu}}, \bibinfo {author} {\bibfnamefont {L.-H.}\ \bibnamefont {Zhang}},
  \bibinfo {author} {\bibfnamefont {Z.-K.}\ \bibnamefont {Liu}}, \bibinfo
  {author} {\bibfnamefont {Z.-Y.}\ \bibnamefont {Zhang}}, \bibinfo {author}
  {\bibfnamefont {Z.-H.}\ \bibnamefont {Zhu}}, \bibinfo {author} {\bibfnamefont
  {W.}~\bibnamefont {Gao}}, \bibinfo {author} {\bibfnamefont {G.-C.}\
  \bibnamefont {Guo}}, \bibinfo {author} {\bibfnamefont {D.-S.}\ \bibnamefont
  {Ding}},\ and\ \bibinfo {author} {\bibfnamefont {B.-S.}\ \bibnamefont
  {Shi}},\ }\bibfield  {title} {\bibinfo {title} {Highly sensitive measurement
  of a megahertz rf electric field with a {R}ydberg-atom sensor},\ }\href
  {https://doi.org/10.1103/PhysRevApplied.18.014045} {\bibfield  {journal}
  {\bibinfo  {journal} {Phys. Rev. Appl.}\ }\textbf {\bibinfo {volume} {18}},\
  \bibinfo {pages} {014045} (\bibinfo {year} {2022}{\natexlab{b}})}\BibitemShut
  {NoStop}%
\bibitem [{\citenamefont {Liu}\ \emph {et~al.}(2022{\natexlab{c}})\citenamefont
  {Liu}, \citenamefont {Zhang}, \citenamefont {Liu}, \citenamefont {Zhang},
  \citenamefont {Guo}, \citenamefont {Ding},\ and\ \citenamefont
  {Shi}}]{liu2022deep}%
  \BibitemOpen
  \bibfield  {author} {\bibinfo {author} {\bibfnamefont {Z.-K.}\ \bibnamefont
  {Liu}}, \bibinfo {author} {\bibfnamefont {L.-H.}\ \bibnamefont {Zhang}},
  \bibinfo {author} {\bibfnamefont {B.}~\bibnamefont {Liu}}, \bibinfo {author}
  {\bibfnamefont {Z.-Y.}\ \bibnamefont {Zhang}}, \bibinfo {author}
  {\bibfnamefont {G.-C.}\ \bibnamefont {Guo}}, \bibinfo {author} {\bibfnamefont
  {D.-S.}\ \bibnamefont {Ding}},\ and\ \bibinfo {author} {\bibfnamefont
  {B.-S.}\ \bibnamefont {Shi}},\ }\bibfield  {title} {\bibinfo {title} {Deep
  learning enhanced {R}ydberg multifrequency microwave recognition},\ }\href
  {https://www.nature.com/articles/s41467-022-29686-7} {\bibfield  {journal}
  {\bibinfo  {journal} {Nat. Commun.}\ }\textbf {\bibinfo {volume} {13}},\
  \bibinfo {pages} {1997} (\bibinfo {year} {2022}{\natexlab{c}})}\BibitemShut
  {NoStop}%
\bibitem [{\citenamefont {Zhang}\ \emph {et~al.}(2022)\citenamefont {Zhang},
  \citenamefont {Liu}, \citenamefont {Liu}, \citenamefont {Zhang},
  \citenamefont {Guo}, \citenamefont {Ding},\ and\ \citenamefont
  {Shi}}]{PhysRevApplied.18.014033}%
  \BibitemOpen
  \bibfield  {author} {\bibinfo {author} {\bibfnamefont {L.-H.}\ \bibnamefont
  {Zhang}}, \bibinfo {author} {\bibfnamefont {Z.-K.}\ \bibnamefont {Liu}},
  \bibinfo {author} {\bibfnamefont {B.}~\bibnamefont {Liu}}, \bibinfo {author}
  {\bibfnamefont {Z.-Y.}\ \bibnamefont {Zhang}}, \bibinfo {author}
  {\bibfnamefont {G.-C.}\ \bibnamefont {Guo}}, \bibinfo {author} {\bibfnamefont
  {D.-S.}\ \bibnamefont {Ding}},\ and\ \bibinfo {author} {\bibfnamefont
  {B.-S.}\ \bibnamefont {Shi}},\ }\bibfield  {title} {\bibinfo {title} {Rydberg
  microwave-frequency-comb spectrometer},\ }\href
  {https://doi.org/10.1103/PhysRevApplied.18.014033} {\bibfield  {journal}
  {\bibinfo  {journal} {Phys. Rev. Appl.}\ }\textbf {\bibinfo {volume} {18}},\
  \bibinfo {pages} {014033} (\bibinfo {year} {2022})}\BibitemShut {NoStop}%
\bibitem [{\citenamefont {Anderson}\ \emph {et~al.}(2020)\citenamefont
  {Anderson}, \citenamefont {Sapiro},\ and\ \citenamefont {Raithel}}]{63}%
  \BibitemOpen
  \bibfield  {author} {\bibinfo {author} {\bibfnamefont {D.~A.}\ \bibnamefont
  {Anderson}}, \bibinfo {author} {\bibfnamefont {R.~E.}\ \bibnamefont
  {Sapiro}},\ and\ \bibinfo {author} {\bibfnamefont {G.}~\bibnamefont
  {Raithel}},\ }\bibfield  {title} {\bibinfo {title} {Rydberg atoms for
  radio-frequency communications and sensing: Atomic receivers for pulsed rf
  field and phase detection},\ }\href
  {https://doi.org/10.1109/maes.2019.2960922} {\bibfield  {journal} {\bibinfo
  {journal} {IEEE Aerospace and Electronic Systems Magazine}\ }\textbf
  {\bibinfo {volume} {35}},\ \bibinfo {pages} {48} (\bibinfo {year}
  {2020})}\BibitemShut {NoStop}%
\bibitem [{\citenamefont {Meyer}\ \emph {et~al.}(2018)\citenamefont {Meyer},
  \citenamefont {Cox}, \citenamefont {Fatemi},\ and\ \citenamefont
  {Kunz}}]{42}%
  \BibitemOpen
  \bibfield  {author} {\bibinfo {author} {\bibfnamefont {D.~H.}\ \bibnamefont
  {Meyer}}, \bibinfo {author} {\bibfnamefont {K.~C.}\ \bibnamefont {Cox}},
  \bibinfo {author} {\bibfnamefont {F.~K.}\ \bibnamefont {Fatemi}},\ and\
  \bibinfo {author} {\bibfnamefont {P.~D.}\ \bibnamefont {Kunz}},\ }\bibfield
  {title} {\bibinfo {title} {{Digital communication with Rydberg atoms and
  amplitude-modulated microwave fields}},\ }\href
  {https://doi.org/10.1063/1.5028357} {\bibfield  {journal} {\bibinfo
  {journal} {Applied Physics Letters}\ }\textbf {\bibinfo {volume} {112}},\
  \bibinfo {pages} {211108} (\bibinfo {year} {2018})}\BibitemShut {NoStop}%
\bibitem [{\citenamefont {Cao}\ \emph {et~al.}(2015)\citenamefont {Cao},
  \citenamefont {Cheung},\ and\ \citenamefont {Yuk}}]{7005405}%
  \BibitemOpen
  \bibfield  {author} {\bibinfo {author} {\bibfnamefont {Y.~F.}\ \bibnamefont
  {Cao}}, \bibinfo {author} {\bibfnamefont {S.~W.}\ \bibnamefont {Cheung}},\
  and\ \bibinfo {author} {\bibfnamefont {T.~I.}\ \bibnamefont {Yuk}},\
  }\bibfield  {title} {\bibinfo {title} {A multiband slot antenna for
  gps/wimax/wlan systems},\ }\href {https://doi.org/10.1109/TAP.2015.2389219}
  {\bibfield  {journal} {\bibinfo  {journal} {IEEE Transactions on Antennas and
  Propagation}\ }\textbf {\bibinfo {volume} {63}},\ \bibinfo {pages} {952}
  (\bibinfo {year} {2015})}\BibitemShut {NoStop}%
\bibitem [{\citenamefont {Morgan}\ and\ \citenamefont
  {Hogan}(2020)}]{PhysRevLett.124.193604}%
  \BibitemOpen
  \bibfield  {author} {\bibinfo {author} {\bibfnamefont {A.~A.}\ \bibnamefont
  {Morgan}}\ and\ \bibinfo {author} {\bibfnamefont {S.~D.}\ \bibnamefont
  {Hogan}},\ }\bibfield  {title} {\bibinfo {title} {Coupling {R}ydberg atoms to
  microwave fields in a superconducting coplanar waveguide resonator},\ }\href
  {https://doi.org/10.1103/PhysRevLett.124.193604} {\bibfield  {journal}
  {\bibinfo  {journal} {Phys. Rev. Lett.}\ }\textbf {\bibinfo {volume} {124}},\
  \bibinfo {pages} {193604} (\bibinfo {year} {2020})}\BibitemShut {NoStop}%
\bibitem [{\citenamefont {Hogan}\ \emph {et~al.}(2012)\citenamefont {Hogan},
  \citenamefont {Agner}, \citenamefont {Merkt}, \citenamefont {Thiele},
  \citenamefont {Filipp},\ and\ \citenamefont
  {Wallraff}}]{PhysRevLett.108.063004}%
  \BibitemOpen
  \bibfield  {author} {\bibinfo {author} {\bibfnamefont {S.~D.}\ \bibnamefont
  {Hogan}}, \bibinfo {author} {\bibfnamefont {J.~A.}\ \bibnamefont {Agner}},
  \bibinfo {author} {\bibfnamefont {F.}~\bibnamefont {Merkt}}, \bibinfo
  {author} {\bibfnamefont {T.}~\bibnamefont {Thiele}}, \bibinfo {author}
  {\bibfnamefont {S.}~\bibnamefont {Filipp}},\ and\ \bibinfo {author}
  {\bibfnamefont {A.}~\bibnamefont {Wallraff}},\ }\bibfield  {title} {\bibinfo
  {title} {Driving {Rydberg}-{Rydberg} transitions from a coplanar microwave
  waveguide},\ }\href {https://doi.org/10.1103/PhysRevLett.108.063004}
  {\bibfield  {journal} {\bibinfo  {journal} {Phys. Rev. Lett.}\ }\textbf
  {\bibinfo {volume} {108}},\ \bibinfo {pages} {063004} (\bibinfo {year}
  {2012})}\BibitemShut {NoStop}%
\bibitem [{\citenamefont {Walker}\ \emph {et~al.}(2022)\citenamefont {Walker},
  \citenamefont {Brown},\ and\ \citenamefont {Hogan}}]{PhysRevA.105.022626}%
  \BibitemOpen
  \bibfield  {author} {\bibinfo {author} {\bibfnamefont {D.~M.}\ \bibnamefont
  {Walker}}, \bibinfo {author} {\bibfnamefont {L.~L.}\ \bibnamefont {Brown}},\
  and\ \bibinfo {author} {\bibfnamefont {S.~D.}\ \bibnamefont {Hogan}},\
  }\bibfield  {title} {\bibinfo {title} {Electrometry of a single resonator
  mode at a {Rydberg}-atom--superconducting-circuit interface},\ }\href
  {https://doi.org/10.1103/PhysRevA.105.022626} {\bibfield  {journal} {\bibinfo
   {journal} {Phys. Rev. A}\ }\textbf {\bibinfo {volume} {105}},\ \bibinfo
  {pages} {022626} (\bibinfo {year} {2022})}\BibitemShut {NoStop}%
\bibitem [{\citenamefont {Holloway}\ \emph {et~al.}(2022)\citenamefont
  {Holloway}, \citenamefont {Prajapati}, \citenamefont {Artusio-Glimpse},
  \citenamefont {Berweger}, \citenamefont {Simons}, \citenamefont {Kasahara},
  \citenamefont {Al{\`u}},\ and\ \citenamefont {Ziolkowski}}]{holloway2022ssr}%
  \BibitemOpen
  \bibfield  {author} {\bibinfo {author} {\bibfnamefont {C.~L.}\ \bibnamefont
  {Holloway}}, \bibinfo {author} {\bibfnamefont {N.}~\bibnamefont {Prajapati}},
  \bibinfo {author} {\bibfnamefont {A.~B.}\ \bibnamefont {Artusio-Glimpse}},
  \bibinfo {author} {\bibfnamefont {S.}~\bibnamefont {Berweger}}, \bibinfo
  {author} {\bibfnamefont {M.~T.}\ \bibnamefont {Simons}}, \bibinfo {author}
  {\bibfnamefont {Y.}~\bibnamefont {Kasahara}}, \bibinfo {author}
  {\bibfnamefont {A.}~\bibnamefont {Al{\`u}}},\ and\ \bibinfo {author}
  {\bibfnamefont {R.~W.}\ \bibnamefont {Ziolkowski}},\ }\bibfield  {title}
  {\bibinfo {title} {{R}ydberg atom-based field sensing enhancement using a
  split-ring resonator},\ }\href {https://doi.org/10.1063/5.0088532} {\bibfield
   {journal} {\bibinfo  {journal} {Applied Physics Letters}\ }\textbf {\bibinfo
  {volume} {120}},\ \bibinfo {pages} {204001} (\bibinfo {year}
  {2022})}\BibitemShut {NoStop}%
\bibitem [{\citenamefont {Yang}\ \emph {et~al.}(2023)\citenamefont {Yang},
  \citenamefont {Mao}, \citenamefont {He}, \citenamefont {Yao}, \citenamefont
  {Li}, \citenamefont {Sun},\ and\ \citenamefont {Fu}}]{RN3337}%
  \BibitemOpen
  \bibfield  {author} {\bibinfo {author} {\bibfnamefont {K.}~\bibnamefont
  {Yang}}, \bibinfo {author} {\bibfnamefont {R.}~\bibnamefont {Mao}}, \bibinfo
  {author} {\bibfnamefont {L.}~\bibnamefont {He}}, \bibinfo {author}
  {\bibfnamefont {J.}~\bibnamefont {Yao}}, \bibinfo {author} {\bibfnamefont
  {J.}~\bibnamefont {Li}}, \bibinfo {author} {\bibfnamefont {Z.}~\bibnamefont
  {Sun}},\ and\ \bibinfo {author} {\bibfnamefont {Y.}~\bibnamefont {Fu}},\
  }\bibfield  {title} {\bibinfo {title} {Local oscillator port embedded field
  enhancement resonator for {Rydberg} atomic heterodyne technique},\ }\href
  {https://doi.org/10.1140/epjqt/s40507-023-00179-w} {\bibfield  {journal}
  {\bibinfo  {journal} {EPJ Quantum Technology}\ }\textbf {\bibinfo {volume}
  {10}},\ \bibinfo {pages} {23} (\bibinfo {year} {2023})}\BibitemShut {NoStop}%
\bibitem [{\citenamefont {Thaicharoen}\ \emph {et~al.}(2019)\citenamefont
  {Thaicharoen}, \citenamefont {Moore}, \citenamefont {Anderson}, \citenamefont
  {Powel}, \citenamefont {Peterson},\ and\ \citenamefont
  {Raithel}}]{PhysRevA.100.063427}%
  \BibitemOpen
  \bibfield  {author} {\bibinfo {author} {\bibfnamefont {N.}~\bibnamefont
  {Thaicharoen}}, \bibinfo {author} {\bibfnamefont {K.~R.}\ \bibnamefont
  {Moore}}, \bibinfo {author} {\bibfnamefont {D.~A.}\ \bibnamefont {Anderson}},
  \bibinfo {author} {\bibfnamefont {R.~C.}\ \bibnamefont {Powel}}, \bibinfo
  {author} {\bibfnamefont {E.}~\bibnamefont {Peterson}},\ and\ \bibinfo
  {author} {\bibfnamefont {G.}~\bibnamefont {Raithel}},\ }\bibfield  {title}
  {\bibinfo {title} {Electromagnetically induced transparency, absorption, and
  microwave-field sensing in a {Rb} vapor cell with a three-color all-infrared
  laser system},\ }\href {https://doi.org/10.1103/PhysRevA.100.063427}
  {\bibfield  {journal} {\bibinfo  {journal} {Phys. Rev. A}\ }\textbf {\bibinfo
  {volume} {100}},\ \bibinfo {pages} {063427} (\bibinfo {year}
  {2019})}\BibitemShut {NoStop}%
\bibitem [{\citenamefont {Santamaria-Botello}\ \emph
  {et~al.}(2022)\citenamefont {Santamaria-Botello}, \citenamefont {Verploegh},
  \citenamefont {Bottomley},\ and\ \citenamefont
  {Popovic}}]{santamariabotello2022comparison}%
  \BibitemOpen
  \bibfield  {author} {\bibinfo {author} {\bibfnamefont {G.}~\bibnamefont
  {Santamaria-Botello}}, \bibinfo {author} {\bibfnamefont {S.}~\bibnamefont
  {Verploegh}}, \bibinfo {author} {\bibfnamefont {E.}~\bibnamefont
  {Bottomley}},\ and\ \bibinfo {author} {\bibfnamefont {Z.}~\bibnamefont
  {Popovic}},\ }\href@noop {} {\bibinfo {title} {Comparison of noise
  temperature of {Rydberg}-atom and electronic microwave receivers}} (\bibinfo
  {year} {2022}),\ \Eprint {https://arxiv.org/abs/2209.00908} {arXiv:2209.00908
  [quant-ph]} \BibitemShut {NoStop}%
\bibitem [{\citenamefont {Kuhr}\ \emph {et~al.}(2007)\citenamefont {Kuhr},
  \citenamefont {Gleyzes}, \citenamefont {Guerlin}, \citenamefont {Bernu},
  \citenamefont {Hoff}, \citenamefont {Deléglise}, \citenamefont {Osnaghi},
  \citenamefont {Brune}, \citenamefont {Raimond}, \citenamefont {Haroche},
  \citenamefont {Jacques}, \citenamefont {Bosland},\ and\ \citenamefont
  {Visentin}}]{10.1063/1.2724816}%
  \BibitemOpen
  \bibfield  {author} {\bibinfo {author} {\bibfnamefont {S.}~\bibnamefont
  {Kuhr}}, \bibinfo {author} {\bibfnamefont {S.}~\bibnamefont {Gleyzes}},
  \bibinfo {author} {\bibfnamefont {C.}~\bibnamefont {Guerlin}}, \bibinfo
  {author} {\bibfnamefont {J.}~\bibnamefont {Bernu}}, \bibinfo {author}
  {\bibfnamefont {U.~B.}\ \bibnamefont {Hoff}}, \bibinfo {author}
  {\bibfnamefont {S.}~\bibnamefont {Deléglise}}, \bibinfo {author}
  {\bibfnamefont {S.}~\bibnamefont {Osnaghi}}, \bibinfo {author} {\bibfnamefont
  {M.}~\bibnamefont {Brune}}, \bibinfo {author} {\bibfnamefont {J.-M.}\
  \bibnamefont {Raimond}}, \bibinfo {author} {\bibfnamefont {S.}~\bibnamefont
  {Haroche}}, \bibinfo {author} {\bibfnamefont {E.}~\bibnamefont {Jacques}},
  \bibinfo {author} {\bibfnamefont {P.}~\bibnamefont {Bosland}},\ and\ \bibinfo
  {author} {\bibfnamefont {B.}~\bibnamefont {Visentin}},\ }\bibfield  {title}
  {\bibinfo {title} {{Ultrahigh finesse Fabry-Pérot superconducting
  resonator}},\ }\href {https://doi.org/10.1063/1.2724816} {\bibfield
  {journal} {\bibinfo  {journal} {Applied Physics Letters}\ }\textbf {\bibinfo
  {volume} {90}},\ \bibinfo {pages} {164101} (\bibinfo {year}
  {2007})}\BibitemShut {NoStop}%
\bibitem [{\citenamefont {Suleymanzade}\ \emph {et~al.}(2020)\citenamefont
  {Suleymanzade}, \citenamefont {Anferov}, \citenamefont {Stone}, \citenamefont
  {Naik}, \citenamefont {Oriani}, \citenamefont {Simon},\ and\ \citenamefont
  {Schuster}}]{10.1063/1.5137900}%
  \BibitemOpen
  \bibfield  {author} {\bibinfo {author} {\bibfnamefont {A.}~\bibnamefont
  {Suleymanzade}}, \bibinfo {author} {\bibfnamefont {A.}~\bibnamefont
  {Anferov}}, \bibinfo {author} {\bibfnamefont {M.}~\bibnamefont {Stone}},
  \bibinfo {author} {\bibfnamefont {R.~K.}\ \bibnamefont {Naik}}, \bibinfo
  {author} {\bibfnamefont {A.}~\bibnamefont {Oriani}}, \bibinfo {author}
  {\bibfnamefont {J.}~\bibnamefont {Simon}},\ and\ \bibinfo {author}
  {\bibfnamefont {D.}~\bibnamefont {Schuster}},\ }\bibfield  {title} {\bibinfo
  {title} {{A tunable high-Q millimeter wave cavity for hybrid circuit and
  cavity QED experiments}},\ }\href {https://doi.org/10.1063/1.5137900}
  {\bibfield  {journal} {\bibinfo  {journal} {Applied Physics Letters}\
  }\textbf {\bibinfo {volume} {116}},\ \bibinfo {pages} {104001} (\bibinfo
  {year} {2020})}\BibitemShut {NoStop}%
\bibitem [{\citenamefont {Gu\'e}\ \emph {et~al.}(2023)\citenamefont {Gu\'e},
  \citenamefont {Hees}, \citenamefont {Lodewyck}, \citenamefont {Le~Targat},\
  and\ \citenamefont {Wolf}}]{PhysRevD.108.035042}%
  \BibitemOpen
  \bibfield  {author} {\bibinfo {author} {\bibfnamefont {J.}~\bibnamefont
  {Gu\'e}}, \bibinfo {author} {\bibfnamefont {A.}~\bibnamefont {Hees}},
  \bibinfo {author} {\bibfnamefont {J.}~\bibnamefont {Lodewyck}}, \bibinfo
  {author} {\bibfnamefont {R.}~\bibnamefont {Le~Targat}},\ and\ \bibinfo
  {author} {\bibfnamefont {P.}~\bibnamefont {Wolf}},\ }\bibfield  {title}
  {\bibinfo {title} {Search for vector dark matter in microwave cavities with
  {Rydberg} atoms},\ }\href {https://doi.org/10.1103/PhysRevD.108.035042}
  {\bibfield  {journal} {\bibinfo  {journal} {Phys. Rev. D}\ }\textbf {\bibinfo
  {volume} {108}},\ \bibinfo {pages} {035042} (\bibinfo {year}
  {2023})}\BibitemShut {NoStop}%
\bibitem [{\citenamefont {Bradley}\ \emph {et~al.}(2003)\citenamefont
  {Bradley}, \citenamefont {Clarke}, \citenamefont {Kinion}, \citenamefont
  {Rosenberg}, \citenamefont {van Bibber}, \citenamefont {Matsuki},
  \citenamefont {M\"uck},\ and\ \citenamefont {Sikivie}}]{RevModPhys.75.777}%
  \BibitemOpen
  \bibfield  {author} {\bibinfo {author} {\bibfnamefont {R.}~\bibnamefont
  {Bradley}}, \bibinfo {author} {\bibfnamefont {J.}~\bibnamefont {Clarke}},
  \bibinfo {author} {\bibfnamefont {D.}~\bibnamefont {Kinion}}, \bibinfo
  {author} {\bibfnamefont {L.~J.}\ \bibnamefont {Rosenberg}}, \bibinfo {author}
  {\bibfnamefont {K.}~\bibnamefont {van Bibber}}, \bibinfo {author}
  {\bibfnamefont {S.}~\bibnamefont {Matsuki}}, \bibinfo {author} {\bibfnamefont
  {M.}~\bibnamefont {M\"uck}},\ and\ \bibinfo {author} {\bibfnamefont
  {P.}~\bibnamefont {Sikivie}},\ }\bibfield  {title} {\bibinfo {title}
  {Microwave cavity searches for dark-matter axions},\ }\href
  {https://doi.org/10.1103/RevModPhys.75.777} {\bibfield  {journal} {\bibinfo
  {journal} {Rev. Mod. Phys.}\ }\textbf {\bibinfo {volume} {75}},\ \bibinfo
  {pages} {777} (\bibinfo {year} {2003})}\BibitemShut {NoStop}%
\bibitem [{\citenamefont {Engelhardt}\ \emph {et~al.}(2023)\citenamefont
  {Engelhardt}, \citenamefont {Bhoonah},\ and\ \citenamefont
  {Liu}}]{engelhardt2023detecting}%
  \BibitemOpen
  \bibfield  {author} {\bibinfo {author} {\bibfnamefont {G.}~\bibnamefont
  {Engelhardt}}, \bibinfo {author} {\bibfnamefont {A.}~\bibnamefont
  {Bhoonah}},\ and\ \bibinfo {author} {\bibfnamefont {W.~V.}\ \bibnamefont
  {Liu}},\ }\href@noop {} {\bibinfo {title} {Detecting axion dark matter with
  {Rydberg} atoms via induced electric dipole transitions}} (\bibinfo {year}
  {2023}),\ \Eprint {https://arxiv.org/abs/2304.05863} {arXiv:2304.05863
  [hep-ph]} \BibitemShut {NoStop}%
\end{thebibliography}%
 
 \end{document}